# Performance Fault Detection in Wind Turbines by Dynamic Reference State Estimation

Angela Meyer[1*] and Bernhard Brodbeck[2]

[1] School of Engineering, Zurich University of Applied Sciences, 8400 Winterthur, Switzerland
[2] WinJi AG, Am Wasser 55, 8049 Zurich, Switzerland

*Corresponding author (angela.meyer@zhaw.ch)

*Abstract*— The operation and maintenance costs of wind parks make up a major fraction of a park's overall lifetime costs. They also include opportunity costs of lost revenue from avoidable power generation underperformance. We present a machine-learning based decision support method that minimizes these opportunity costs. By analyzing the stream of telemetry sensor data from the turbine operation, estimating highly accurate power reference relations and benchmarking, we can detect performance-related operational faults in a turbine- and site-specific manner. The most accurate power reference model is selected based on combinations of machine learning algorithms and regressor sets. Operating personal can be alerted if a normal operating state boundary is exceeded. We demonstrate the performance fault detection method in a case study for a commercial grid-connected onshore wind turbine. Diagnosing a detected underperformance event, we find that the observed power generation deficiencies coincide with rotor blade misalignment related to low hydraulic pressure of the turbine's blade actuators.

## I. INTRODUCTION

THE globally installed wind power capacity reached 591 GW at the end of 2018 and continued growth is expected by at least 55 GW annually until 2023 [1]. The operation and maintenance (O&M) costs of commercial wind turbines constitute a large fraction of the overall lifecycle costs. They can amount to 30% and more of the lifecycle costs of onshore and offshore wind parks [2], [3]. As the purchase prices of wind turbines continue to decrease while O&M cost remain relatively constant, O&M activities are becoming more and more important as a relative cost factor [3]. High reliability and availability and low O&M costs are important economic requirements for wind plants in view of the projected strong investment increase in the globally installed wind power capacity.

While the weather conditions are the main determinant of the wind turbine power generation, optimal technical conditions are important as well for the cost-effective operation and thus need to be monitored. A variety of condition monitoring approaches have been proposed for wind turbines, including oil analysis, vibration, acoustics and strain monitoring [4]. More recently several multivariate data mining approaches to condition monitoring have been proposed for detecting performance-related faults and quantifying underperformance in the power generation of wind parks. These approaches make use of turbine telemetry data comprising sensor-based and control variables for estimating power curve models in a parametric or non-parametric manner [5]. Several parametric methods have been proposed, e.g. [6]–[8]. Among the parametric approaches, reference [6] proposed a monitoring method which calculates a power curve by spline interpolation of mean power values per wind speed bin and then learns upper and lower limits to the power curve and issues an alarm if the limits are exceeded. Being based on discrete bins and parametric interpolation, this method may be prone to provide less accurate estimates than highly flexible machine learning regression models. Moreover, it issues alarms based on single 10-minute average outlier values, which may result in a large number of alarms even in case of brief transient deviations. An underperformance detection in the power generation was proposed in [7] based on power curve modelling with stepwise linear models and Weibull cumulative distribution functions. It is found that such parametric approaches can be limited in their flexibility to capture power relations and to reflect the characteristics of individual turbines and sites.

Among the non-parametric approaches, several regression, clustering and dimensionality-reduction based methods have been proposed, e.g. [9]–[12]. Reference [9] introduced a monitoring method for wind parks that is based on a non-parametric estimation approach and three operational curves: power, rotor speed and pitch angle as a function of the wind speed. The curves were segmented into clusters with a k-means algorithm to enable the detection of outliers based on the Mahalanobis distance from the curves. A comparison to the reference operational curves was performed only indirectly by comparing the skewness and tailedness of the multivariate power, rotor speed and pitch angle distributions. While the statistical moments used, skewness and kurtosis, characterize

A. Meyer is associate professor at the School of Engineering of Zurich University of Applied Sciences, 8400 Winterthur, Switzerland. *A. Meyer is the corresponding author (e-mail: angela.meyer@zhaw.ch, phone: +41 58 934 67 74).

B. Brodbeck is the founder of the WinJi AG, Am Wasser 55, 8049 Zurich, Switzerland (e-mail: bernhard.brodbeck@win-ji.com). This work was partially supported by the Swiss innovation agency Innosuisse Grant 31533.1.

the overall shape of the multivariate distribution they do not enable a unique identification of the reference distribution. So a highly accurate detection of deviations from the reference distribution may not always be possible with this approach. Moreover, the operation reference relations were not estimated in a turbine or site-specific manner even though said reference relations can exhibit systematic differences across turbines and sites, even for turbines of the same type and in the same wind park. It is still a challenge to accurately estimate the expected power generation and detect power-related operational faults. The more accurately the operational power reference relation can be estimated, the more reliably one can detect power generation underperformance situations and perform diagnosis. An accurate estimation of the wind power relationship permits the operators to quantify the potential power generation and assess the wind power loss and, if needed, decide on appropriate measures with regard to calibration or aerodynamics and control system optimization and maintenance.

The goal of this study is to develop a highly accurate machine-learning based decision support method for detecting and quantifying power generation anomalies in wind turbines in an automated turbine- and site-specific manner. To achieve this, we propose a method that consists of estimating reference wind power relations for each individual turbine under normal operation conditions in a highly accurate manner, followed by selection of the most accurate power relation representation derived from a variety of combinations of machine learning algorithms and regressor sets, and then detection of significant deviations from the expected turbine output. We demonstrate this approach to power generation-related fault detection in the case of a commercial grid-connected onshore wind turbine.

## II. METHODOLOGY

### A. Turbine and dataset

This study makes use of one year of historical wind turbine operation data from the supervisory control and data acquisition (SCADA) system of a commercial 3.3 MW rated power turbine. We anonymized the turbine data to protect the operator's privacy. Hence the location and observation time periods are not disclosed. The wind turbine is a commercial horizontal-axis type with three blades positioned upwind of the tower following a standard commercial design with drive train with gear box and asynchronous electrical generator positioned in the nacelle at the top of the tower. The turbine is equipped with various sensing devices, SCADA data collection and condition monitoring systems. It carries environmental sensors including an anemometer and a thermometer to measure the wind speed, direction and the air temperature at the nacelle. The latest SCADA data is provided every 10 minutes with values averaged over 10-minute intervals. The data comprises system state variables including nacelle yaw angles, blade pitch angles, hydraulic pressure and oil temperatures, gear bearing temperatures, rotor and generator speeds, voltages, currents and generated power along with environmental condition variables including wind speed, wind direction and air temperature. The turbine's cut-in and cut-out wind speeds are at around 3 m/s and 25 m/s, respectively.

Three-bladed horizontal-axis wind turbines of several MW rated power make up the large majority of commercial wind turbines in operation today. The studied turbine is an upwind variable-speed model in which the rotor speed is adjusted to achieve the optimal tip speed ratio and extract the maximum power from the wind at any given wind speed above the cut-in and below the rated wind speeds. The turbine power control is achieved through variable rotor speed control and pitch regulation. Below the rated wind speed, the blade angle control unit pitches each blade along its axis in response to changes in the wind speed. This is done in order to achieve an optimal angle of attack of the incoming wind so as to maximize the power output at all wind speeds between the cut-in and the rated wind speed. The rotor hub carrying the blades is attached to the nacelle at the top of the turbine tower. Each rotor blade is mounted at the hub via a blade bearing that facilitates the orientation of the blades at a given pitch angle. This enables control of the loads that act on the blades by setting the blades' orientation with respect to the wind. The attack angle of the wind on the blades is set by the pitch control unit in conjunction with a pitch cylinder for each blade. The pitch control determines the power generated by the turbine by controlling the blade pitch angle so as to optimize the turbine's efficiency. The pitch angle is increased at winds speeds above the rated speed in order to reduce the load acting on the blades, the drive train and generator and prevent damage to them.

### B. Algorithm

The present method does not rely on any requirements with regard to the turbine site or the availability of additional meteorological measurement equipment as required for power curve calibration measurements or warranty-related power performance assessments [13], [14]. We propose to detect power generation underperformance in wind turbines in an operational manner through an accurate estimation of the power generation potential and continuous monitoring for and identification of any significant deviations from the estimated power generation. We quantify the power generation potential based on a multidimensional wind power model that is estimated from SCADA data taken under normal operating conditions. The relationship of the generated power and its driving factors is non-linear and limited at the rated power of the turbine which is 3.3 MW in this study. The wind speed to power relationship $P \sim v_{\text{wind}}$ is characterized by a cut-in velocity below which the turbine is not generating but consuming power from the electrical grid. As the wind speed increases, the turbine's rated power production level may be reached. From this point onwards, the turbine power generation will be controlled to not exceed the rated power. Turbine control systems are designed to maximize the fraction of wind kinetic energy conversion into electrical energy between the cut-in wind speed and the rated wind speed. The rotor blade pitch angle and rotor speed are controlled to achieve this goal by maximum power point tracking. The power reference relationship thus depends on the specifics of the individual turbine design and control. It can also be affected by the characteristics of the site at which the turbine was built. We address this challenge by estimating a model of the power generation potential based on a set of most important explanatory variables, including the wind speed, the wind direction and the air temperature, $T_{air}$. The latter is used as an imperfect proxy of the air pressure $p_{air}$ in accordance with the ideal gas law $p_{air} = \rho_{air} R T_{air}$ wherein $\rho_{air}$ and $R$ are the air density and the specific gas constant. Generally speaking, we propose to estimate the expected power based on the most suitable set of regressor variables combined with a statistical or machine learning model that provides the most accurate estimate.

For the case of the 3.3 MW turbine of this study, the power generation potential is estimated based on three different sets of explanatory variables: the wind speed $v_{wind}$ as the only regressor variable, the wind speed and wind direction $\alpha_{wind}$, and the wind speed, direction and air temperature $T_{air}$ measured at the nacelle,

$$P \sim v_{wind} \tag{1}$$

$$P \sim v_{wind} + \alpha_{wind} \tag{2}$$

$$P \sim v_{wind} + \alpha_{wind} + T_{air}. \tag{3}$$

A variety of machine learning and statistical algorithms are adopted for estimating the model parameters for all three relations: a Random Forest and a Gradient Boosting Machine (GBM) algorithm, Support Vector Machines (SVM), k Nearest Neighbours (kNN), a Bayesian Regularized Neural Network (BRNN) algorithm and two parametric approaches for comparison: a Generalized Additive Model (GAM) with locally estimated smoothing (LOESS) and a GAM without local smoothing and with stepwise feature selection based on the Akaike information criterion (AIC). The stepwise feature selection is particularly relevant the larger the set of regressors.

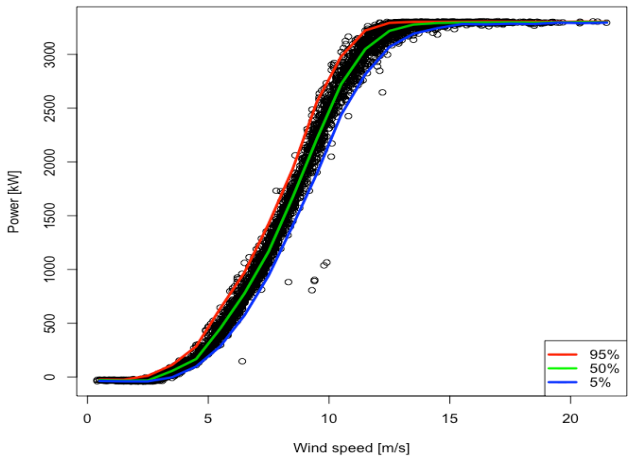

Fig. 1. Power curve estimated based on SCADA data of the reference month prior to removal of outliers. The 5% and 95% power quantiles and the median are indicated for the wind speed bins.

The models estimated for the case study that we present below are based on historical SCADA data from one month of wind turbine operation (named "reference month" herein after) when the turbine was in normal fault-free operation. In this time period the weather onsite was characterized by a wide range of conditions, with air temperatures ranging from -1 to 16 °C and wind speeds between 0.4 and 21.5 m/s. This enabled us to estimate the full power relationship based on a relatively representative range of regressor values. The estimated model parameters are specific to the particular turbine and site. The explanatory variables provided to the algorithms are 10-minute averages based on measurements by the anemometer installed at nacelle height of the turbine and the air temperature at nacelle height. The target variable to be estimated is the 10-minute average net power $P$ generated by the wind turbine. A randomized train/test split was performed wherein 70% of the data is used for estimating the model parameters and the remaining 30% is used for quantifying the accuracy of the fitted models. The estimated power reference relation can then be used to estimate the expected power generation based on recent measurements of wind speed, direction and air temperature.

As part of the data preprocessing, time periods were the SCADA system logged warnings or errors were removed along with the data immediately preceding or following these time periods. Further automated data cleaning was performed by removing outliers based on the estimated power reference relation and then re-estimating the power reference relation as explained before. Outliers are data points that are at a relatively large distance from most other observations. They were removed from the training set to achieve a more accurate power reference relation. This was done by estimating upper and lower confidence bands around the reference relation by segregating the wind speed domain into disjoint wind speed bins of integer wind speed values of [0 m/s, 1 m/s), [1 m/s, 2 m/s), and so forth. Upper and lower quantiles, specifically 5% and 95% quantiles of generated power, were computed for each wind speed bin. The resulting set of upper and lower power quantiles is shown in Fig. 1. Outliers were defined as points lying outside of the 5% to 95% power quantile range by more than half of the interquantile distance for the purpose of estimating the reference power relation. They were removed from the training and test sets for the estimation of the reference relation.

III. RESULTS

The prediction accuracies based on the estimated power relations are shown in Table 1 in terms of the root mean square error (RMSE, in kilowatts) and the coefficient of determination $R^2$ for different statistical and machine learning algorithms. The highest accuracies with RMSE < 76 kW can be achieved with the random forest, GBM, BRNN and SVM algorithms estimating $P \sim v_{wind} + \alpha_{wind} + T_{air}$. The $R^2$ values of up to 0.997 are in line with the high achieved estimation accuracies. We propose to automatically select the most accurate model of the power reference relation according to the lowest RMSE. For the turbine investigated in this study, the random forest algorithm achieves the most accurate predictions (estimates) with all three regressor variables: wind speed, wind direction and air temperature. However, it required a much longer training time than the GBM algorithm which provides comparable prediction accuracy, so we present the case study based on the power reference model estimated with the GBM algorithm. Moreover, we found that the automated data cleaning based on wind speed bins significantly improved the estimation accuracy.

TABLE I
ESTIMATION ACCURACIES FOR $P \sim V_{WIND} + \alpha_{WIND} + T_{AIR}$

| Algorithm | RMSE in kW | $R^2$ |
|---|---|---|
| Random Forest | 67.1 | 0.997 |
| Gradient Boosting | 68.1 | 0.997 |
| BRNN | 72.8 | 0.996 |
| SVM | 75.5 | 0.996 |
| kNN | 185.5 | 0.977 |
| GAM | 360.4 | 0.913 |
| GAM LOESS | 136.8 | 0.987 |

Power reference relation estimation accuracies based on different machine learning algorithms.

Our goal in this context is to develop a method for detecting power generation anomalies in individual wind turbines in a turbine and site-specific manner and diagnose potential causes. We achieve this by applying the trained model as a reference of the turbine's power generation capacity under normal operation conditions and detecting deviations from the expected power output in a moving window approach. The GBM algorithm achieves a root mean square error of 68 kW when trained on the three regressors. The relation between the estimated and actually delivered power is illustrated in Fig. 2 which shows that the trained power reference model provides unbiased estimates at a high prediction accuracy of 2% of the turbine's rated power. The area above the diagonal line in the left-hand panel of Fig. 2 shows multiple outlier values in the actually generated power indicating cases of turbine underperformance.

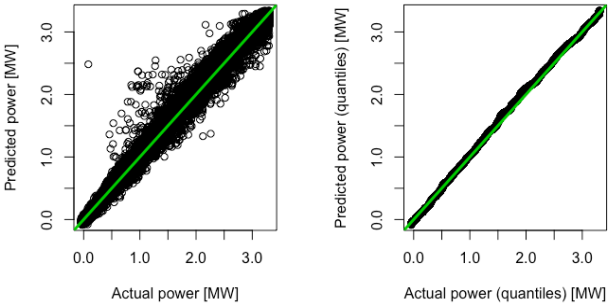

Fig. 2. Power generation of the turbine and corresponding power estimates in MW based on the GBM model for the entire year. A quantile-quantile representation is shown in the right-hand panel.

The energy that the turbine was expected to deliver over the past 24 hours is computed according to the estimated power reference model and the observed environmental conditions at each point in time. The expected and the delivered energy can each also be aggregated and monitored over continuous time intervals other than 24 hours, for instance 2 or 30 hours. The shorter a time horizon is chosen, the more sensitive the metrics will be with regard to detecting short-term power generation deficiencies. The residual between the actually delivered and the expected energy is computed over the past 24 hours in our case. The distribution of the residuals and the evolution of the residuals over the course of the year is shown in Fig. 3 for the turbine investigated in this study.

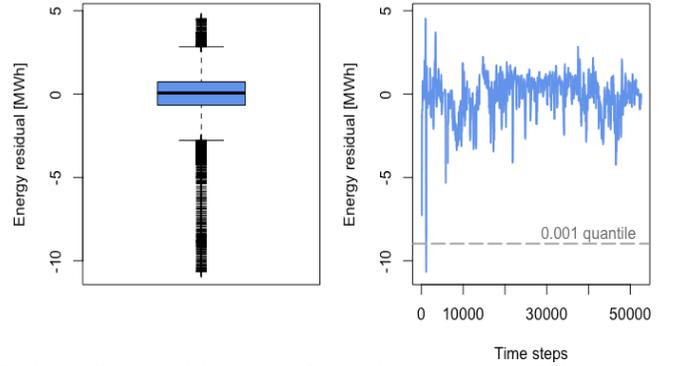

Fig. 3. Distribution and time series of residual energy over past 24 hours over the course of the year.

Various other performance metrics were assessed in this study with regard to their suitability for detecting power generation anomalies, including:

$$M_1 = P - P_{exp} \qquad (4)$$

$$M_2 = |P - P_{exp}| \qquad (5)$$

$$M_3 = (P - P_{exp})/P_{exp} \qquad (6)$$

$$M_4 = |(P - P_{exp})/P_{exp}| \qquad (7)$$

$$M_5 = P/P_{exp} \qquad (8)$$

$$M_6 = E/E_{exp} \qquad (9)$$

wherein $P$ and $P_{exp}$, as well as $E$ and $E_{exp}$ denote the actual and the expected power and energy provided by the turbine, respectively. These metrics were found to be not optimal for the task at hand. For instance, $M_5$ is the ratio of the actual power delivered over the expected power generation. $M_5$ can be affected by the uncertainty in the shape and position of the reference power relation, in other words the variance in the regressor variables wind speed, wind direction and air temperature, and the actual power generation. Uncertainty in the wind speed values can originate from an anemometer measurement, for instance, which can be affected by sensor biases, electronic noise and response latency. In a similar manner the generated power also has an associated uncertainty. Moreover, the net power provided can be negative at wind speeds below the cut-in velocity because the turbine does not provide power but consumes electricity from the electrical grid in this regime to maintain the turbine control electronics operative. These circumstances can even render $M_5$ negative. Moreover, small $M_5$ values do not necessarily coincide with events where the actual power production significantly falls short of the expected production in absolute terms. $M_5$ also picks up on events barely relevant in absolute power. On the other hand, $M_1$ and $M_2$ are sensitive to large power anomalies in single observations, which is generally too fast for any maintenance technicians to respond to and thus often not of interest in practice. At the same time smaller but systematic deviations on the time scale of hours or days can go undetected with these metrics even though such deviations would be of interest to turbine operators. We refer to [15] for further discussion of alternative power generation metrics for wind turbines.

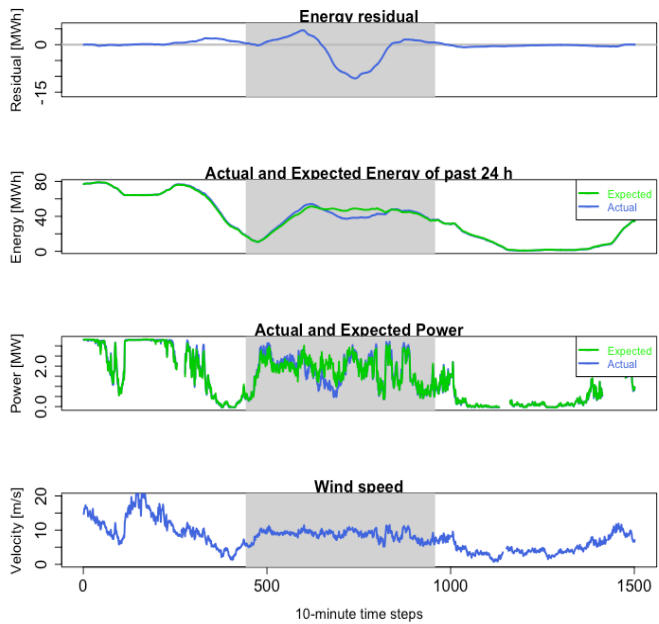

Fig. 4. Actual and expected energy generation in the week of the event. One time step corresponds to 10 minutes.

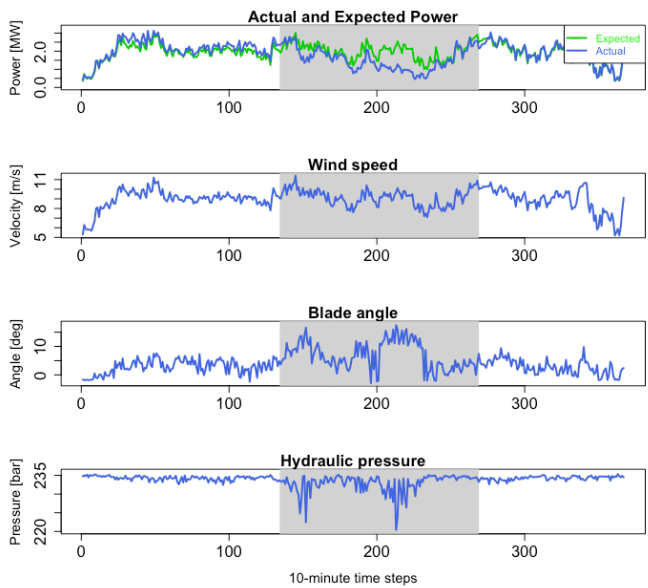

Fig. 5. The power underperformance event is evident over three adjacent days in the beginning of the turbine operation year. The actual power generated falls short of about 1 MW during the event (top panel) even though the wind speed remains relatively constant at around 9 m/s. The power deficiency coincides with hydraulic pressure loss and unusual large blade pitch angles.

In the present study power deficiencies are detected based on the residual between the actually delivered and the expected energy, $E - E_{\exp}$, over the past 24 hours. This metric indicates the degree of economic loss in terms of opportunity cost incurred from turbine operational deficiencies. The presented method is applied to detect events of large wind energy losses. For automated monitoring and alerting, a threshold value can be applied which can be derived based on quantiles of the residual distribution, for instance in normal operation regime. For the turbine in this study, only 0.1% of all energy residuals were larger in absolute value than 8.9 MWh, which constitutes the 0.001 quantile of all residuals observed over the course of the year investigated in this study. The largest residual of the turbine energy generation within 24 hours amounted to 10.3 MWh in absolute value and occurred in the beginning of the measurement period in the vicinity of time step 1000 shown in Fig. 3 right-hand panel. Comparing the actual and the expected energy generation in Fig. 4, it is evident that an underperformance in power generation occurred for about 15 hours during this incident. Up to half of the expected power generation was lost in this time period. Assuming a typical wind farm in Europe, one such event per turbine per year can translate to annual opportunity cost on the order of 20'000 EUR. Hence such power deficiencies affect the turbine profitability, amongst others, and may require the attention of the asset management and O&M teams if persistent.

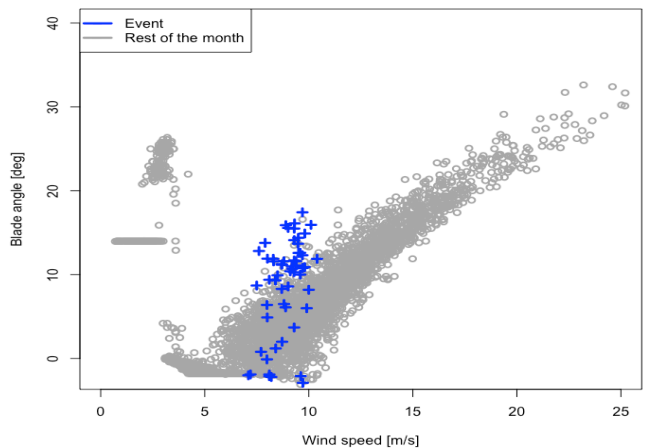

Fig. 6 Pitch angles of the turbine blades during the power underperformance event compared to the rest of the month in which the event occurred.

Fig. 5 illustrates the power deficiency event in more detail. Assessing the turbine operational data prior to and during the event as shown in Figs. 5 and 6, it is evident that the expected power and the actual power match with high accuracy before and after the event. The average wind speed was relatively constant during the event, ranging between 7 and 10 m/s. At the same time particularly low hydraulic pressure values were measured and the rotor blades were pitched at larger-than-usual angles at the given wind speeds (Fig. 6). This suggests that the power generation deficiencies observed during the incident have been caused by a rotor blade misalignment related to low hydraulic pressure of the rotor blade actuators. The hydraulic system is typically a turbine subsystem with relatively high annual failure rates [16]. Indeed, we found that in the case of the studied turbine the hydraulic pressure control system responsible for the blade pitching was updated five months after the event and fewer significant power deficiencies were observed after the update (Fig. 3).

IV. CONCLUSION

The operation and maintenance costs of wind turbines constitute a major fraction of their lifetime costs. We have presented a machine-learning based approach for detecting performance-impacting operational faults in wind turbines that enables site- and turbine-specific underperformance detection. Our approach can be applied to SCADA data generated during the turbine operation. It continuously analyzes the stream of operational data to detect anomalous operating states and does not require a physics-based wind-turbine model nor additional measurement equipment because it relies on the accurate estimation of the expected power generated under the environmental and operational conditions at the turbine site in a normal operational state. Turbine-type and site-specific differences are accounted for. The developed method can provide decision support to turbine operators and maintenance personal. We have demonstrated it in a case study for a commercial grid-connected onshore wind turbine in which we focus on time periods where the turbine is running in normal fault-free operation according to the automatic control system information provided by the manufacturer. A significant power generation underperformance event was identified that lasted for about 15 hours in which, according to the turbine control system, the turbine was in a normal fault-free operational state. Diagnosing the event, we found that the observed power generation deficiencies coincided with rotor blade misalignment and low hydraulic pressure of the turbine's rotor blade actuators. We expect our approach to aid at identifying power production anomalies and their root causes in order to provide maintenance decision support to turbine operators, thus facilitating the high-performance cost-effective operation of wind parks.

ACKNOWLEDGMENTS

The authors would like to thank Innosuisse, the Swiss innovation agency, for their support of this project. They would also like to thank Christoph Heitz, Janine Maron, Linda Boedi and Dimitrios Anagnostos for discussions. A.M. designed the methodology, analysed the data, created the figures and wrote the draft manuscript. B.B. collected and preprocessed the data, provided the business context, contributed with discussions and to the final manuscript.

REFERENCES

[1] Global Wind Energy Council, 2019, News release: 51.3 GW of global wind capacity installed in 2018, accessible: https://gwec.net/51-3-gw-of-global-wind-capacity-installed-in-2018/ (accessed on 27 March 2020)
[2] Karyotakis, A., 2011, On the Optimisation of Operation and Maintenance Strategies for Offshore Wind Farms, PhD Dissertation, Department of Mechanical Engineering, University College London
[3] Metcalfe, O., 2020, Onshore O&M – New Strategies in a Maturing Sector, Conference presentation at Wind Operations Europe on 5 March 2020
[4] Tchakoua, P., R. Wamkeue, M. Ouhrouche, F. Slaoui-Hasnaoui, T. Tameghe, G. Ekemb, 2014, Wind Turbine Condition Monitoring: State-of-the-Art Review, New Trends, and Future Challenges, *Energies*, https://doi.org/10.3390/en7042595
[5] Lydia, M., S. Kumar, I. Selvakumar, E. Kumar, 2014, A comprehensive review on wind turbine power curve modeling techniques, *Renewable and Sustainable Energy Reviews*, https://doi.org/10.1016/j.rser.2013.10.030
[6] Park, J., J. Lee, K. Oh, J. Lee, 2014, Development of a Novel Power Curve Monitoring Method for Wind Turbines and Its Field Tests, *IEEE Transactions on Energy Conversion*, https://doi.org/10.1109/TEC.2013.2294893
[7] Long, H., L. Wang, Z. Zhang, Z. Song, J. Xu, 2015, Data-Driven Wind Turbine Power Generation Performance Monitoring, *IEEE Transactions on Industrial Electronics*, https://doi.org/10.1109/TIE.2015.2447508
[8] Ruiz de la Hermosa Gonzales-Carrato, R., 2018, Wind farm monitoring using Mahalanobis distance and fuzzy clustering, *Renewable Energy*, https://doi.org/10.1016/j.renene.2018.02.097
[9] Kusiak, A., A. Verma, 2013, Monitoring Wind Farms With Performance Curves, *IEEE Transactions on Sustainable Energy*, https://doi.org/10.1109/TSTE.2012.2212470
[10] Pelletier, F., C. Masson, A. Tahan, 2016, Wind turbine power curve modelling using artificial neural network, *Renewable Energy*, https://doi.org/10.1016/j.renene.2015.11.065
[11] Wang, S., Y. Huang, L. Li, C. Liu, 2016, Wind turbines abnormality detection through analysis of wind farm power curves, *Measurement*, https://doi.org/10.1016/j.measurement.2016.07.006
[12] Pandit, R., D. Infield, 2018, SCADA-based wind turbine anomaly detection using Gaussian process models for wind turbine condition monitoring purposes, *IET Renewable Power Generation*, https://doi.org/10.1049/iet-rpg.2018.0156
[13] International Electrotechnical Commission, IEC61400-12-1, 2005, Wind turbines part 12-1: Power performance measurements of electricity producing wind turbines, International Electrotechnical Commission, Geneva
[14] International Electrotechnical Commission, IEC61400-12-2, 2013, Wind turbines Part 12-2: Power performance of electricity-producing wind turbines based on nacelle anemometry, International Electrotechnical Commission, Geneva
[15] Niu, B., H. Hwangbo, L. Zeng, Y. Ding, 2018, Evaluation of alternative power production efficiency metrics for offshore wind turbines and farms, *Renewable Energy*, https://doi.org/10.1016/j.renene.2018.05.050
[16] Ozturk, S., V. Fthenakis, S. Faulstich, 2018, Failure Modes, Effects and Criticality Analysis for Wind Turbines Considering Climatic Regions and Comparing Geared and Direct Drive Wind Turbines, *Energies*, https://doi.org/10.3390/en11092317